\documentclass[10pt,twocolumn]{article} 

\usepackage[utf8]{inputenc} 
\usepackage[english]{babel}

\usepackage{geometry} 
\usepackage{graphicx} 
\usepackage{paralist} 
\usepackage{verbatim} 
\usepackage{subfig} 
\usepackage{lipsum}  
\usepackage{hyperref}
\usepackage[superscript]{cite}  

\usepackage[format=plain,
            labelfont=it,
            textfont=it]{caption}

\usepackage{fancyhdr} 
\usepackage{varwidth}
\usepackage{authblk}
\newcommand{\filiacion}[2]{\affil[#1]{\protect\begin{varwidth}[t]{\linewidth}\protect\centering \normalfont#2 \protect\end{varwidth}}}

\newcommand{\corresponding}[2]{\author[#1]{\bf #2\thanks{}}}
\newcommand\cauthemail[1]{\footnotetext{#1}}
\newcommand{\fecha}[1]{\date{\vspace{-1ex}\small{#1}}}
\newcommand{\titulo}[2]{\title{\bf{\large{#1 \\ \vspace{1.5ex} #2 }}}}

\newcommand{\enresumen}[1]{\small{#1 \par}\vspace{1.5ex}}
\newcommand{\keywords}[1]{\small{\emph{#1} \par}\vspace{1.5ex}}

\usepackage{sectsty}
\allsectionsfont{\fontsize{10}{12}\sffamily\bfseries\upshape} 

\usepackage{titlesec}
\titlespacing*{\section}{0pt}{1.5ex}{0.8ex}
\titlespacing*{\subsection}{0pt}{1.2ex}{0.6ex}
\usepackage[nottoc,notlof,notlot]{tocbibind} 
\usepackage[titles,subfigure]{tocloft} 

\usepackage{booktabs} 
\usepackage{array} 
\makeatletter
\newcommand{\thickhline}{%
    \noalign {\ifnum 0=`}\fi \hrule height 1.5pt
    \futurelet \reserved@a \@xhline
}
\newcolumntype{"}{@{\hskip\tabcolsep\vrule width 1pt\hskip\tabcolsep}}
\makeatother
\newcolumntype{L}[1]{>{\raggedright\let\newline\\\arraybackslash\hspace{0pt}}m{#1}}
\newcolumntype{C}[1]{>{\centering\let\newline\\\arraybackslash\hspace{0pt}}m{#1}}
\newcolumntype{R}[1]{>{\raggedleft\let\newline\\\arraybackslash\hspace{0pt}}m{#1}}

\makeatletter
\renewcommand\@biblabel[1]{#1.}
\makeatother

\titulo{Un método heurístico para obtener la envolvente de señal y su implementacion de software}
{A heuristic approach to obtain signal envelope with a simple software implementation}

\corresponding{1,2}{C. Jarne} 

\filiacion{1}{Universidad Nacional de Quilmes- Departamento de Ciencia y Tecnología}
\filiacion{2}{CONICET}

\fecha{Recibido: xx/xx/xx; Aceptado: xx/xx/xx} 

\begin{document}

\renewcommand{\abstractname}{}
\twocolumn[
  \begin{@twocolumnfalse}
   \maketitle
    \begin{abstract}\vspace{-12ex}
\centering\begin{minipage}{\dimexpr\paperwidth-6cm}


\enresumen{Signal amplitude envelope allows to obtain information of the signal features for different applications. It is widely used to pre-process sound and other signals of physiological origin in human or animal studies. In order to  obtain signal envelope, a fast and simple algorithm is proposed based on peak detection. The procedure presented here is quite straightforward and can be used in different applications of time series analysis. It can be applied in signals with different origin and frequency content. This algorithm is implemented based on python libraries. The open source code is also provided. Aspects on the parameter selection are discussed to adapt the same method for different applications. Also traditional methods are revisited and compared with the one proposed here.}


\end{minipage}
\vspace{4ex}
 \end{abstract}
  \end{@twocolumnfalse}
]

\thispagestyle{empty}

\setcounter{footnote}{1}
\cauthemail{cecilia.jarne@unq.edu.ar}  

\section{Introduction} \label{intro}
In audio signals, envelope is usually defined together with temporal fine structure. The amplitude modulations of animal sounds, speech or musical instruments are important features that often can be used to understand the physics or biology behind different processes. Accurate estimation of the amplitude, or equivalently energy, or envelope of a time-domain signal (waveform) is not trivial. Ideally, the amplitude envelope should outline the waveform connecting the main peaks and avoiding over fitting as was pointed out in \cite{envelope-paper}.  
	
The definition of envelope in the literature is often ambiguous and lacks of an exact mathematical definition. It is commonly agreed that envelope varies slowly and, in some empirical view, it should pass the prominent peaks of the data smoothly \cite{envelope-paper-2}.

Amplitude envelope is a relevant acoustic feature. It allows to get a reliable estimation that follows closely sudden variations in amplitude and avoids ripples in more stable regions. Depending on the fundamental frequency of the signal, it is possible to have a near optimal order selection in some regions. Amplitude variations depending on the application allows to obtain information on the signal patterns or features. i. e. peak detection, maximum, signal sudden variations, etc.


In addition to sound signals, the signal envelope calculation (or similar) is useful in
many application of signal analysis in different fields such as biology, medicine,
geology, material sciences, among others \cite{aplication-01,aplication-02,aplication-03,aplication-04,aplication-05,aplication-06,aplication-07,nuevo}. A simple algorithm together with a code implementation could became a useful tool to be provide for experimental analysis across different disciplines.

Currently to obtain signal envelope there is a diversity of standard methods implemented in different programming languages. Some of the most sophisticated has not been developed in open source code. On the other hand, most of the traditional methods do not allow to obtain a not attenuated envelope. Here a method to obtain the signal envelope is proposed and described. Additionally an open source implementation in python code is provided.

This work is organized as follows: Section \ref{sec-current} is used to provide a review of existing and most popular techniques. These techniques were also implemented with python libraries and compare with the algorithm proposed here. A comment on the more sophisticated algorithms is also presented. The implementation of these methods is not justified in relation to the programming effort versus the gain in processing. Also the fact of these works does not deal with standard methods, whose implementations are not available in open source code, is a point against the philosophy of present work. The algorithms are not straightforward to implement and the analysis is beyond the scope of present work. The main purpose of this works is to propose a simple algorithm to get the amplitude signal envelope in any situation with a direct implementation. In Section \ref{sec-metodo-nuevo} the proposed algorithm is described. The implementation is explained in Section \ref{imple} and some examples of application are provided in Section \ref{par-selec}. This section includes the discussion of how to chose the parameters in order to adapt the method for different signals sources and some cues to lead to good practices in signal analysis are commented. The conclusion are presented in section \ref{conclu}. The code is available at the local University repository. In Section \ref{source} the reference to download the code is provided.

\section{Current envelope techniques}\label{sec-current}

\subsection{Classical techniques}
 
The classical approach for the estimation of the amplitude envelope of a time-domain signal is the technique known as envelope follower \cite{paper-envelope-cita-01,paper-envelope-cita-04}. This method has been widely used and has several implementations across different programming languages and also analog circuits. It consists basically of rectifying the waveform and then low-pass filtering it. This can be implemented in the analog or digital domain. With programming tools, the algorithm consist of taking the absolute value of the digitalized signal, and then use a digital low-pass filter. Several programing languages has libraries with build-in filters such as MATLAB \cite{matlab} or Python \cite{python-01}. For the comparison that is presented in further in Section \ref{sec-metodo-nuevo} we used filters from python libraries.

Another interesting direct approach is to calculate the instantaneous root mean square (RMS) value of the waveform through a sliding window with finite support \cite{paper-envelope-cita-09}. Some authors have proposed other techniques to obtain a more reliable estimation of the amplitude envelope of waveforms. An early attempt \cite{paper-envelope-cita-04} consisted of a piece-wise linear approximation of the waveform. The amplitude envelope is created by finding and connecting the peaks of the waveform in a window that moves through the data.  Python libraries ware used to generate a function to smooth the signal via RMS together with the sliding window and then compare it with the implementation of the method corresponding to the present work (Section \ref{sec-metodo-nuevo}).


Other classical technique consists of using the analytic signal from Hilbert Transform. Notably, if the Hilbert transform of $S(t)$ is equal to its quadrature signal, then the estimates are equal to the actual information signals \cite{paper-envelope-cita-01}. Synthetic signals (i.e. AM) can be constructed to have this property, but there is no reason to expect that animal sounds, acoustic musical instrument sounds or speech also present it. A more realistic condition is verified when we are dealing with narrow-band signals \cite{paper-envelope-cita-14}. In sounds with rich spectral content it is not possible to use this approach because that condition is not fulfilled.

A straightforward approach is also an algorithm that directly use the prominent peaks and finally get the envelope by interpolation. A detail of this method can be found in \cite{envelope-paper-2}. It is an iterative process that needs to add noise to the original signal. In \cite{envelope-paper-2} not open source code is provide. 

\subsection{Regarding more sophisticated techniques}


A sophisticated method developed with not attenuated envelope is called Cepstral Smoothing \cite{envelope-paper}. It uses the real cepstrum defined as the inverse Fourier transform of the log magnitude spectrum. Regarding the log magnitude spectrum of a signal, it is possible to interpret each cepstral coefficient as a measure of the energy present in discrete frequency bands of that signal. Low-pass filtering the cepstrum  would result in a smoother version of the log magnitude spectrum. If we only want to represent the spectral envelope, discarding information about the partials we should set the cutoff frequency below the period of the signal. This method is more sophisticated that the standard techniques and was discussed in detail in \cite{envelope-paper}, but not software implementation has been provided either.

Another method is called empirical mode decomposition (EMD) \cite{envelope-paper-3} allows to obtain the envelope as a result of an optimization precess and sought as a minimum of a quadratic cost function. EMD is an adaptive data analysis method for analyzing nonlinear and non-stationary data, which was proposed by Huang et al. in 1998.
A solution of this optimization problem is obtained in this study, and it is shown that the method could be widely implemented by choosing free parameters, where the frequency resolution or the number of intrinsic mode functions (IMFs) are tuned as well as the shape of the envelopes.

Both method showed to be interesting and accurate. Nevertheless as explain in Section \ref{intro} a disadvantage is that there are not published open source code implementations or libraries of these more sophisticated envelope techniques in any programing language of both works. This prevent us to perform the direct comparison of these algorithms with standard techniques or the one proposed in this paper without having to develop and debugging additional code. 

In following section the algorithm developed for present work is described. 

\section{Current method: an envelope based on a direct peak detection} \label{sec-metodo-nuevo}

To obtain an envelope output that will not be attenuated, a possible approach is to use a peak detection algorithm combined with a moving window. In the approach presented here, every value in the small window bunch is replaced by the maximum. Later it is possible to lowpass-filter the resulting signal afterwards to get rid of the remaining staircase. This process is summarized in the schematic presented in Figure \ref{Fig_0}. In more detail, the method implemented in present work consist of three simple steps. The steps are very basic and allows to obtain a smooth signal following the amplitude variations as precisely as required.

\begin{figure}[htb!]
\hspace*{-0.1cm}\includegraphics[totalheight=9.75cm]{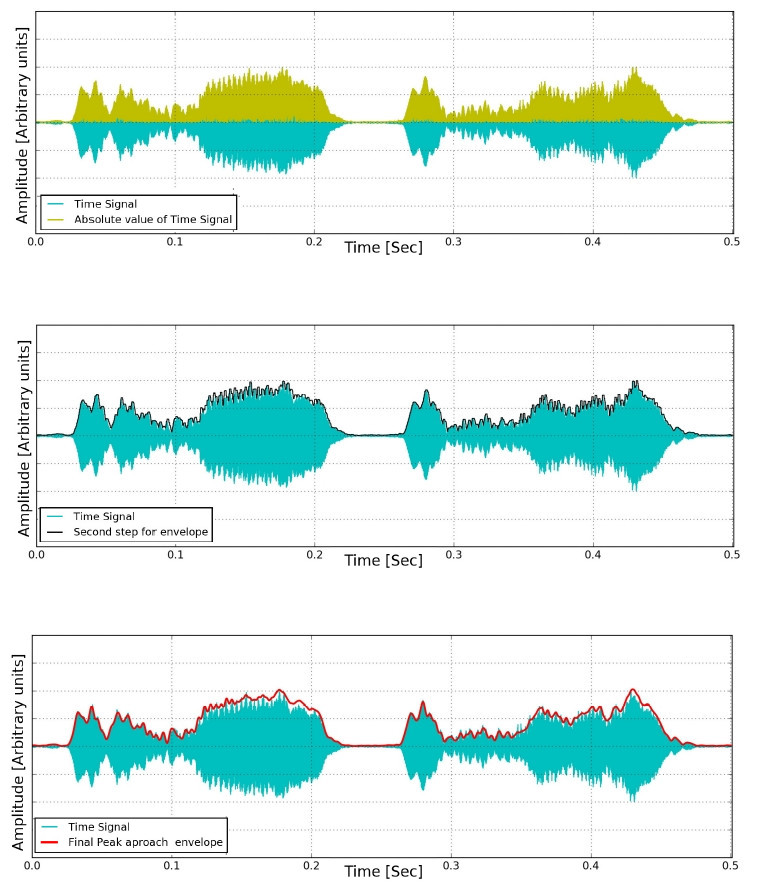}
\caption{Zoom over a song canary segment. Top: Absolute value of time signal. Middle: Peak detection algorithm (pre-envelope). Bottom: envelope, the upper signal when a low pass filter is applied. Filter cutoff frequency is $120$ $KHz$ and window bunch size of $35$ samples.}
\label{DATA-FIG-1}

\end{figure}

\begin{itemize}

\item \textbf{The first step} is to take absolute value of signal $S(t)$, meaning $|S(t)|$ as it is shown in Figure \ref{DATA-FIG-1} top panel. 

\item \textbf{In the second step} it is necessary to divide the $|S(t)|$ in $k$ in non overlapping bunches of $N$ samples:
\vspace{-0.5cm}
\begin{center}
\begin{equation}
|S(t)|=S_1(t)+...+S_i(t)+...+S_k(t),
\end{equation}
\end{center}  

with each $S_i(t)>0$. Then the maximum amplitude value is taken from the signal in each bunch: $Max(|S_i(t)|)=M_i$, corresponding to the $j$ element value in each bunch (peak detection). At this point each value in the bunch is replaced by the maximum $M_i$. In this step the signal $R(t)$ looks like Fig. \ref{DATA-FIG-1} middle panel.
\\
\item \textbf{The third step} consist of a lowpass-filter applied to the resulted signal $R(t)$ to get rid of the remaining staircase ripple to obtain $\hat{S}(t)$ , as illustrated in the  Figure \ref{DATA-FIG-1} bottom panel. 

\end{itemize}

\begin{center}
\vspace{0.5cm}
\begin{figure*}[h]
\begin{center}
\includegraphics[totalheight=3.27cm]{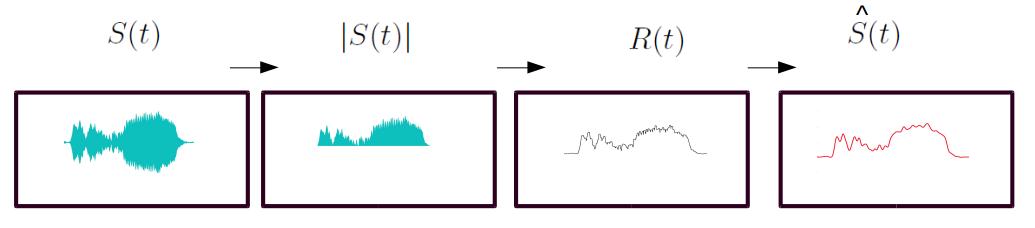}  
\caption{Algorithm schema. Input signal is called $S(t)$ and the envelope obtained is $\hat{S}(t)$. Each arrow represents one step of the algorithm.}
\label{Fig_0}
\end{center}
\end{figure*}
\end{center}

The signal obtained with this technique has the same sample rate that the original signal. This could be very useful in order to compare the envelope with other simultaneous measurements related to the original signal.

By using this three-step method the envelope obtained follows the signal amplitude variations without attenuation, if the parameters are carefully selected as shown in Fig.\ref{DATA-FIG-1}. In this example the filter cutoff frequency is chosen between $100-150$ $Hz$ to ensure that the signal envelope follows signal variations as rapid as $10$ $mS$ combined with the window bunch size of $35$ samples. These two variables could be adjusted to follow variations as rapid as needed, depending on the required signal application.

\begin{figure*}[htb!]
\begin{center}
\hspace*{-1cm}\includegraphics[totalheight=7.5cm]{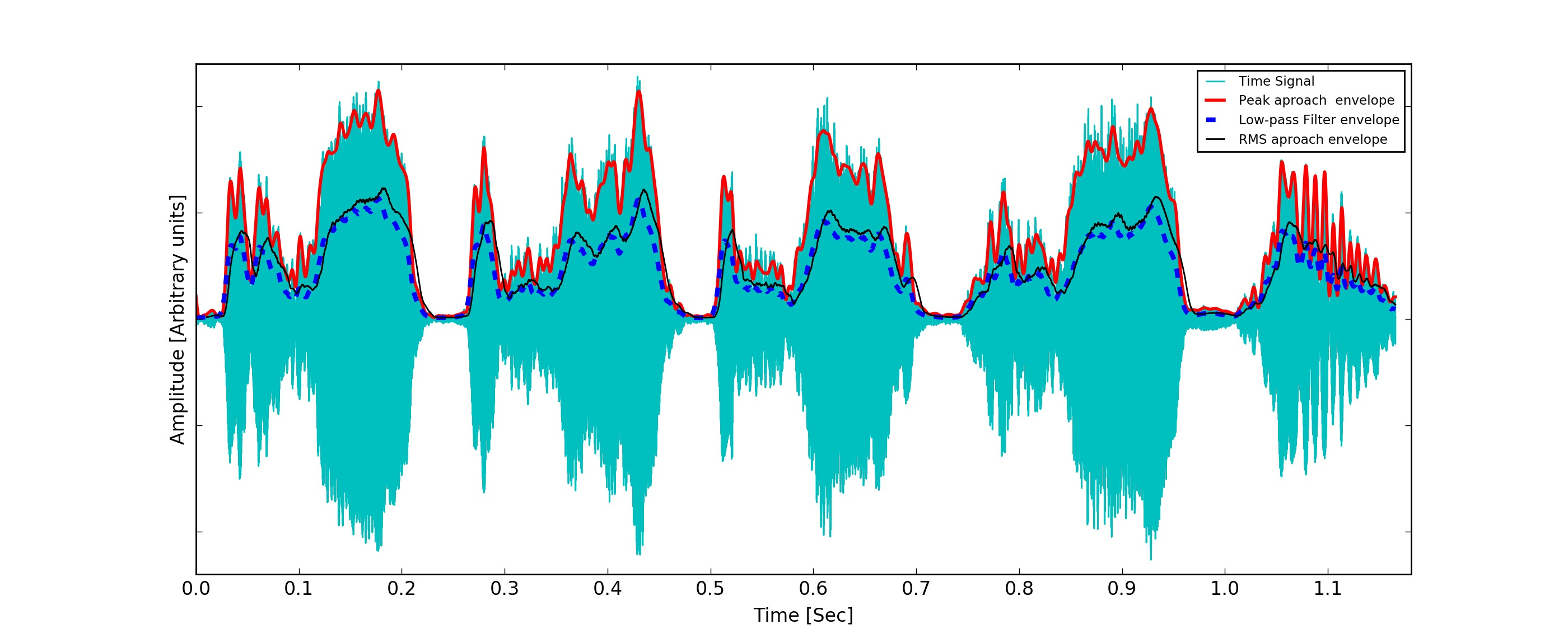}
\caption{Comparison between between the classical implementations of envelope methods and the one proposed in this work. Blue dashed line for low-pass filtered envelope, black thin line corresponds to RMS slide window and red line the proposed method.}
\label{Fig_02}
\end{center}
\end{figure*}

A comparison between the classical implementations of envelope methods (the envelope follower and RMS approach) and the one proposed in this work is presented in Figure \ref{Fig_02}. The code provided in section \ref{source} allows to compared these methods for any audio signal of any sample rate and frequency content.
In the example selected for the figure, the frequency cut for the envelope follower approach in this example is $150$ $Hz$. For the RMS mothod the slide window is 50 samples. For the three-step algorithm parameters are selected as in Figure \ref{DATA-FIG-1}. Main difference between standard methods and the one proposed here is regarding the attenuation.

\section{Discussion and Software implementation} \label{imple}

The previous section was a general description of the method presented. Some particular considerations that has to be taken into account for the code implementation are discussed in this section. In particular the libraries that ware used, the functions and the selection of the parameters in the code, there are also presented as the arguments for the choices that ware made.

The language used for the implementation was Python \cite{python-01}. The main reason to use python is the expansive library of open source data analysis tools, also the web frameworks and the testing instruments that make one of the largest current programming community. Another reason is that Python is an accessible language for new programmers and the community provides many introductory resources.

Regarding the implementation parameters, the size of the bunch for the signal to be divided is an important one. It will mainly depend on the sampling rate $f_S$ but also on the frequency content and time scale variations that the user wants to capture from the original signal. For the examples presented here sample rate is $44.100$ $KHz$ and frequency content is between $100$ $Hz$ and $10000$ $Hz$. In this case a proper bunch size is between $20-200$ samples.To chose larger bunch size means that signal envelope means that the envelope deviates from the amplitude of the original signal but very small bunches means that user will capture small ripple variations of the signal.

Since we have audio files available to analyze, the \textit{Scipy Signal processing library} functions was used to get the audio sample rate and signal vector and write a code with these input variables. 

With respect to the digital filter, it is necessary to choose a filter that does not produce phase shift differences (a zero-phase filtering. It means that does not shift the signal as it filters. Since the phase is zero at all frequencies, it is also linear-phase. Filtering backwards in time requires to predict the future signal value. This kind of filters can not be used for ``online" real-life applications, only for off-line processing of recordings of signals or if we filter signal with some small-delay as it is acquired. 

The \textit{scipy signal library} \cite{python-02} has two kinds of functions to chose for filter design. One is called \textit{scipy.signal.lfilter} and the other \textit{scipy.signal.filtfilt}. The first one (\textit{lfilter}) is causal forward-in-time filtering only, similar to a real-life electronic filter. It cannot be zero-phase. It usually adds different amounts of delay at different frequencies. The appropriated method for the off-line application is to use \textit{filtfilt} with no shift in the signal as it filters.

Once the zero-phase filter is selected, the following step consist of choosing the type of filter as well as the frequency behavior. In this case we used a Butterworth low pass filter type \textit{scipy.signal.butter} of fourth order. Frequency attenuation (behavior) of this filter is elsewhere \cite{filter}. The other key point to obtain the envelope appropriate for the specific application is to choose the frequency cut of the low pass filter from \textbf{Third step} in the algorithm. Depending on this value you can obtain as many detail on the amplitude as it is required.

The method proposed here is not computational expensive, so it can be applied in different systems. For instance for 1.5 second of signal recording with a sample rate of $44.1$ $KHz$, the computational time needed to run the code is $< 0.5$ $Sec$.

To summarize, the algorithm keys for the implementation are: proper bunch size, chose a zero-phase filter,  and chose a frequency cut to follow the signal changes as required.

\section{Results: example of the implementation over different sounds and parameter selection} \label{par-selec}

\begin{figure*}[htb!]
\begin{center}
\hspace*{-0.9cm}\includegraphics[totalheight=6.13cm]{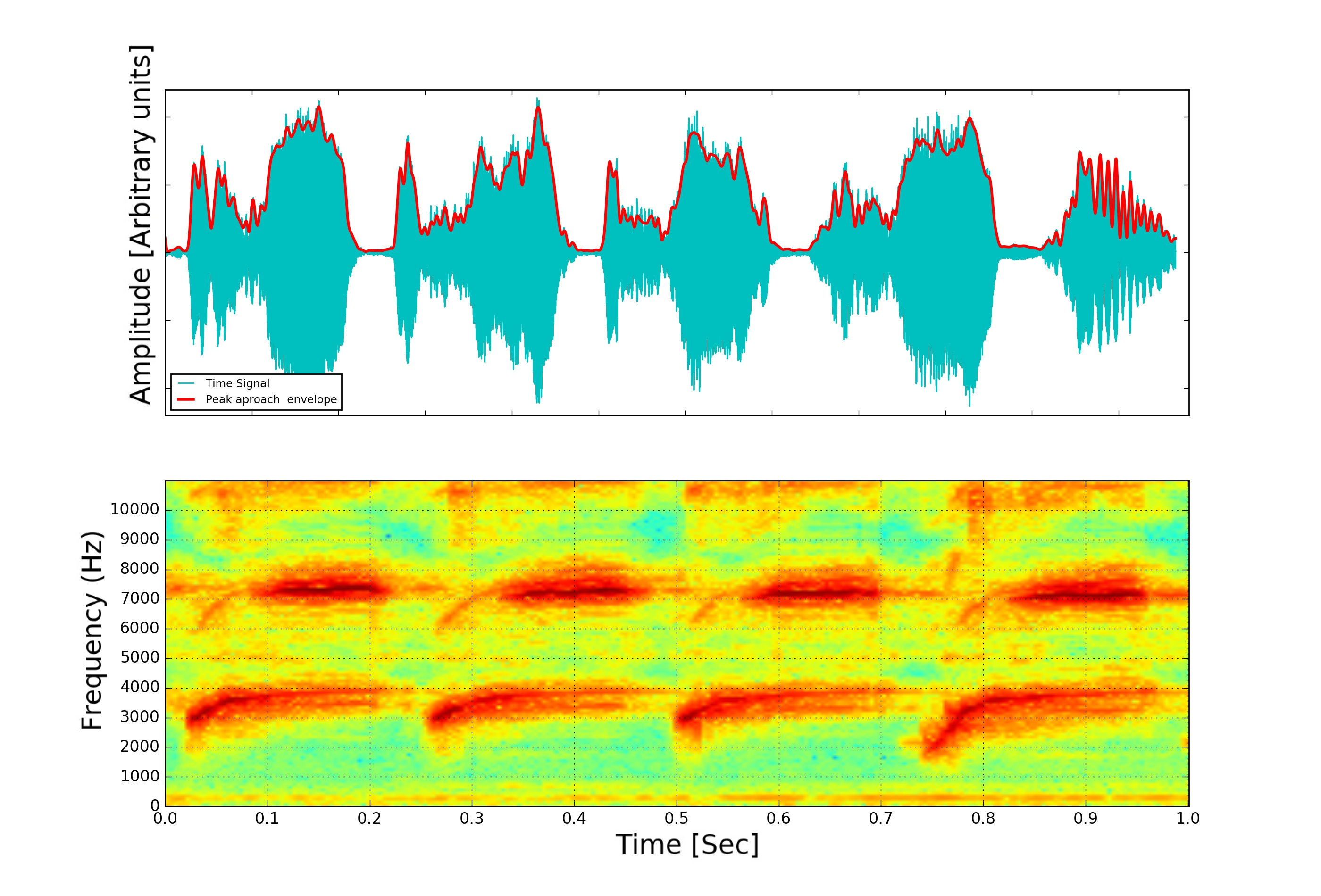}
\hspace*{-0.80cm}\includegraphics[totalheight=6.13cm]{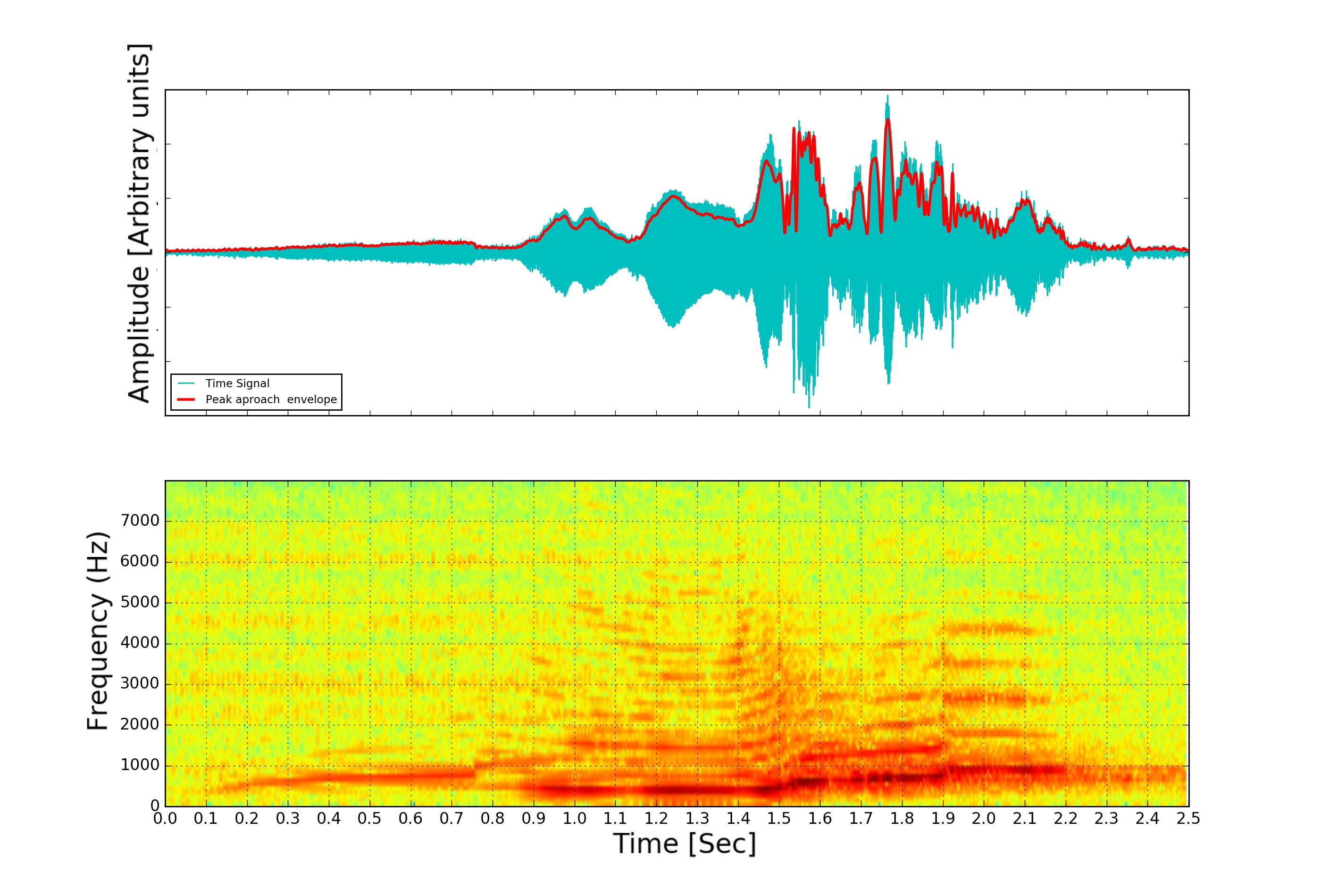}
\hspace*{-0.8cm}\includegraphics[totalheight=6.13cm]{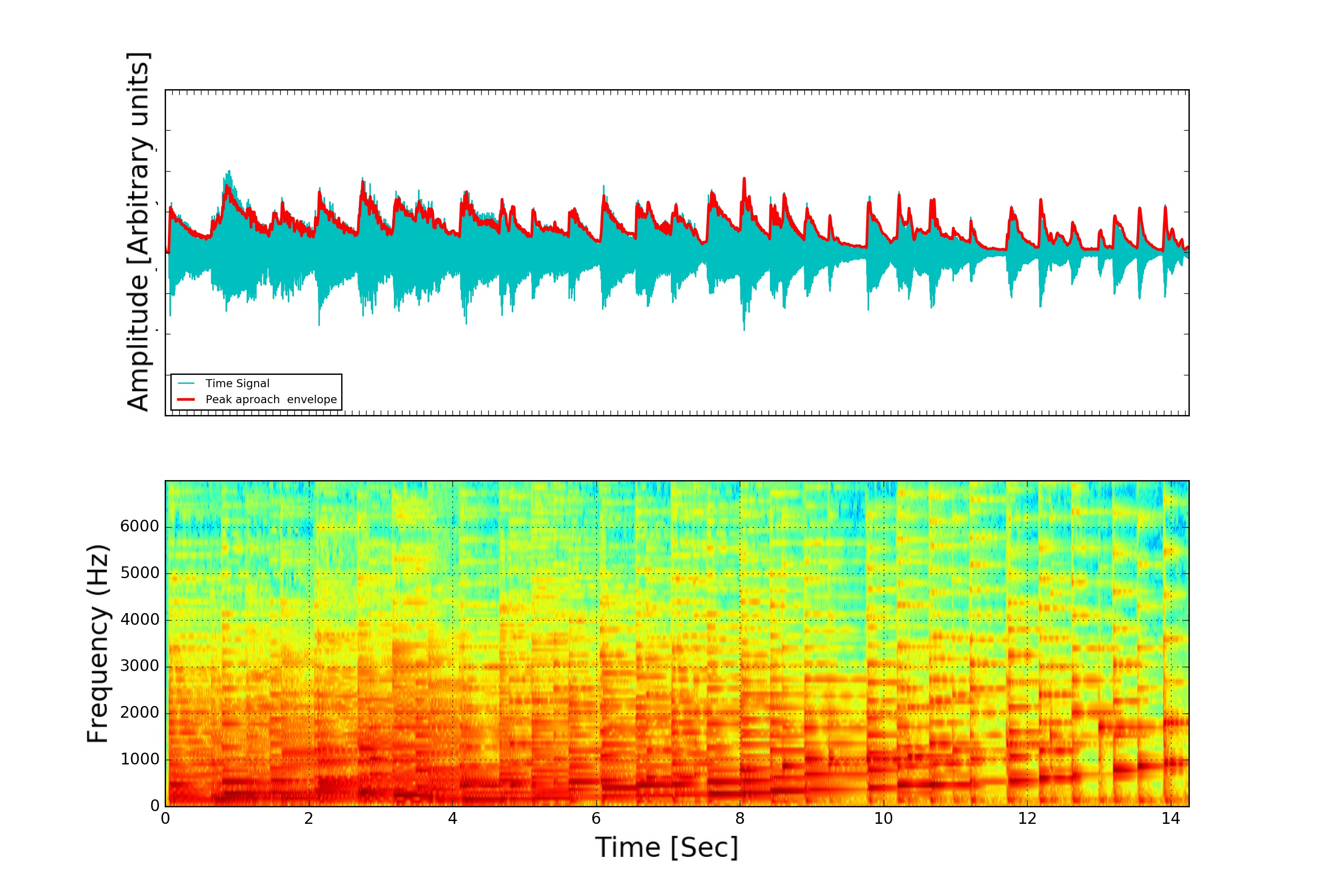}
\hspace*{-0.90cm}\includegraphics[totalheight=6.13cm]{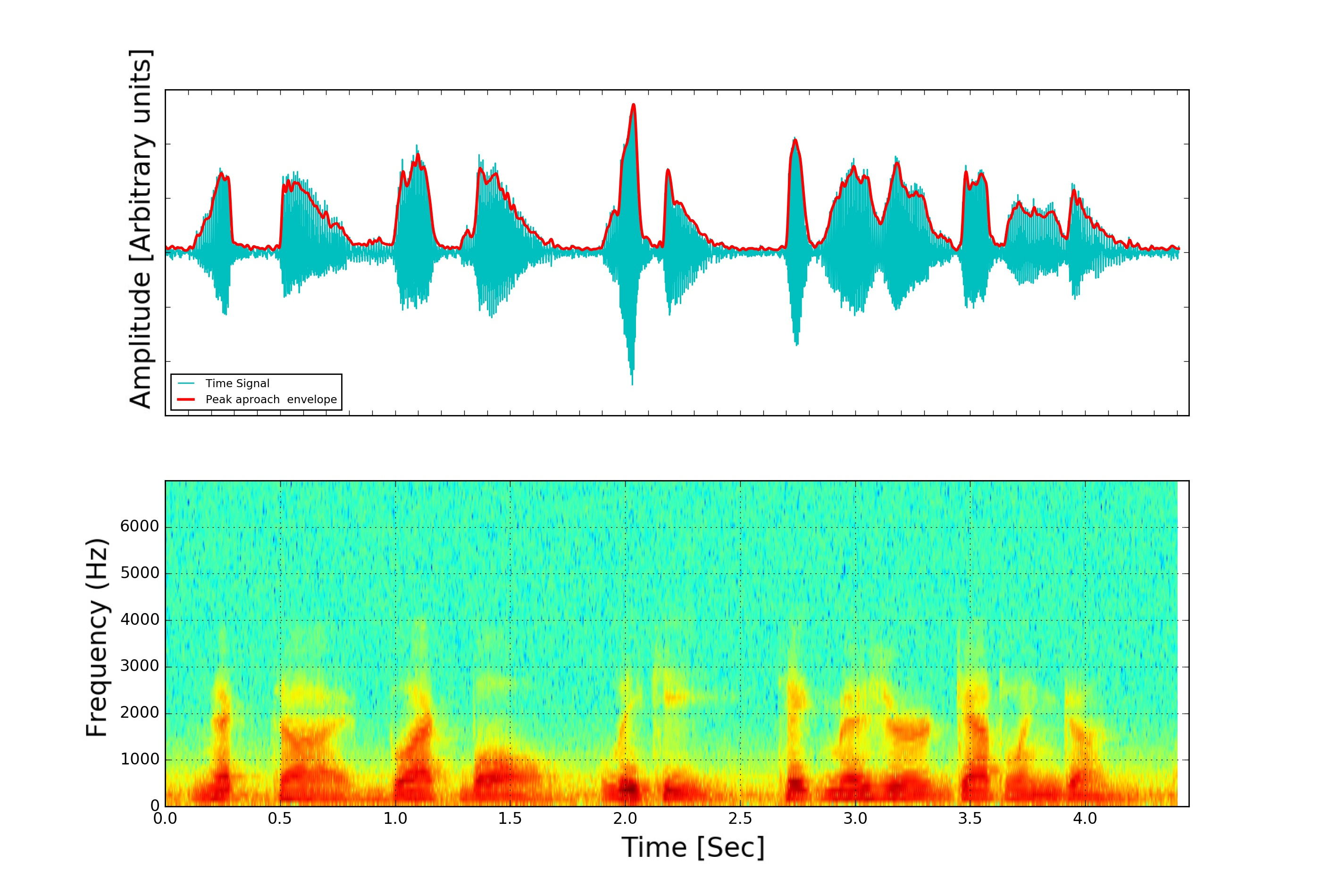}
\vspace*{0.3cm}
\caption{Up Left: Piano song segment where the frequency of the played notes increases over time and the corresponding sonogram. Up Right: Segment of spoken words in English. Down Left: A canary song signal with the estimated envelope and its sonogram. Down Right: A whale song with the corresponding envelope. Frequency content of both signals and also time scales are very different, although the method works in both cases.}
\label{Fig_3}
\end{center}
\end{figure*}

Four different sounds were selected from public data bases to test the algorithm. The envelope result from the method implemented is shown over different kinds of sound segments with or without rapid variations. Audio files was selected from: two sounds of very different animals presented in Figure \ref{Fig_3} (canary song and a whale song), human speech (English language) and an instrument recording of piano song presented with corresponding spectrogram. Table \ref{T_tabla-01} summarizes the parameter values for each case. Sound files and code are provided as supplementary material.

The examples was selected with different time scales as well as frequency content.  The parameter selection will depend on the user needs. A heuristic search for the values convenient in a particular application will depend on how many detail is required to follow the signal variation.

A good method must be flexible enough to be applied to different kinds of signals. In this case the focus was on sound signal, but it is not exclusive. The parameters of the algorithm can be selected depending on the application and the type of signal to be applied.

\begin{tiny}

\begin{table*}[hbt!]

\begin{center}

\begin{tabular}{|c|c|c|}
\hline
\textbf{Source} & \textbf{bunch size{[}samples{]}} & \textbf{Cut Frequency {[}Hz{]}} \\ \hline
Canary sound \cite{tabla_01}          & 35                     & 300                 \\ \hline
Whale sound \cite{tabla_02}           & 50                     & 300                 \\ \hline
Words spoken in English \cite{tabla_03} & 50                   & 100                 \\ \hline
Piano stair scale notes \cite{tabla_04} & 200                  & 100                 \\ \hline
\end{tabular}

\vspace{0.5cm}
\caption{Examples for the values of the parameters selected for samples of different kind of sound origin. Sample rate is $44.1$ $KHz$ for all segments in the table.}
\label{T_tabla-01}
\end{center}
\end{table*}

\end{tiny} 

\vspace*{0.2cm}

\section{Conclusions} \label{conclu}

In this work a heuristic and simple technique is proposed for envelope estimation and a open source code is provided. This algorithm used is very simple to implement and it is not just another empirical method. Even when the procedure is straightforward and simple, its robust solution for different signals and required resolution on signal detail. An important advantage of this method is that it can be applied in signal with different spectral content. The parameters can be adjusted for the time scaling required by the user in order to select the smoothness of the estimated envelope. Experimental applications presented shows that the algorithm can achieve good quality on the temporal envelope estimation of data. It was observed that the estimated envelope can track quite well the changing trend of input data in different time scales. 

This algorithm can be implemented in off-line processing applications, e.g. animal sounds (particularly bird songs), musical instrument sound analysis, heart sound analysis, vibration signal analysis, and etc. In this paper, temporal audio signals were used as examples. However,there are no requirements on the specific domain of the data. It is appropriate for one dimensional data (e.i. time domain, frequency domain, space domain, etc.). Additionally a python implementation is presented to be used with any sound input.

A further key to apply the algorithm in almost real-time implementation, is to process signal time series in segments larger than 1 bunch size. It could be useful depending on the sample rate of the digitalized signal.


\subsection*{Availability of data and source code} \label{source}
The software was developed on python. Code is available under password envelope at University local repository:\\
$http://ceciliajarne.web.unq.edu.ar/investigacion$ and \\
$https://github.com/katejarne/Envelope\_Code$

\section{
\vspace{-3ex}}  
\bibliographystyle{abbrv}
\renewcommand{\bibname}{}
 \bibliography{referencias_anales}

\end{document}